\documentclass[aps,prd,twocolumn,amsmath,amssymb,floatfix, superscriptaddress, longbibliography]{revtex4-1}
\usepackage[dvipsnames]{xcolor}
\usepackage{graphicx}
\usepackage[colorlinks,citecolor=blue,linkcolor=blue,urlcolor=blue,plainpages=false,pdfpagelabels]{hyperref}
\usepackage{bm}

\newcommand{\beq}{\begin{equation}}
\newcommand{\eeq}{\end{equation}}
\newcommand{\bea}{\begin{eqnarray}}
\newcommand{\eea}{\end{eqnarray}}

\begin{document}
\title{Phase transition in quasi-flat band superconductors}

\author{A.~A.~Zyuzin}
\affiliation{---}

\author{A.~Yu.~Zyuzin}
\affiliation{A. F. Ioffe Physical--Technical Institute,~194021 St.~Petersburg, Russia}

\begin{abstract}
We investigate superconductivity in a two-dimensional material described by a two-band heavy-fermion model, where hybridization between a dispersive band and a flat band introduces a quasi-flat dispersion to the otherwise localized flat-band electrons.  The enhanced density of states in the quasi-flat band raises the crossover temperature for an inhomogeneous preformed Cooper pair state. The superconducting phase stiffness and the Berezinskii-Kosterlitz-Thouless (BKT) temperature are governed by the Fermi surface contribution induced by hybridization. We compute the crossover and BKT temperatures, revealing a dome-like dependence on doping. 
When the pairing amplitude exceeds the energy width of the quasi-flat band, superconductivity is suppressed, and the inhomogeneous pairing regime expands linearly with increasing interaction strength. 
However, in the opposite case, the BKT temperature reaches a maximum value that is only numerically less than the energy width of the quasi-flat band. 
We also discuss our results in the context of superconductivity in graphene-based systems.
\end{abstract}

\maketitle
\section{Introduction}
The topic of superconductivity in systems hosting flat-bands and extended van Hove singularities has been studied theoretically for more than three decades, driven by its relevance to strongly correlated electron systems and unconventional superconducting phases \cite{Khodel_Shaginyan, Nozieres, Volovik_1994, Dzyaloshinskii_review, PhysRevLett_Masanori, Physica_C_Furukawa, PhysRevB_Kopnin_Heikkila_Volovik, Yudin_PhysRevLett, Heikkila_twisted, Wu_Macdonald_twisted}. Recent experimental investigations have reported evidence of superconductivity \cite{Exp_TBG_superconductor, Dean_science} accompanied by the signatures of pseudogap \cite{Yazdani_nature} in graphene-based structures with quasi-flat electronic bands promoting the flatronic superconductivity research area; for a review, see \cite{Volovik_review, balents_review}. More recent experiments have also observed the signatures of an anisotropic nodal gap amplitude in twisted bilayer \cite{ Tanaka_2025} and trilayer \cite{Banerjee_2025} graphene.

Quasi-flat bands are characterized by a significantly enhanced density of states, potentially leading to higher temperatures of Cooper pair formation, \cite{Khodel_Shaginyan, Nozieres, Volovik_1994, Dzyaloshinskii_review}. 
In two-dimensional structures, however, achieving global phase coherence requires a stiff superconducting phase, in accordance with the Berezinskii-Kosterlitz-Thouless (BKT) theory, \cite{B, KT, Emery_Kivelson, Nature_Sacepe_Feigelman}.
In such systems, long-range correlations between Cooper pairs, may arise, for example, from finite dispersion of the quasi-flat band itself or hybridization between the quasi-flat band and other coexisting bands in realistic materials.

Recent theoretical studies have highlighted the quantum metric contribution to the phase stiffness \cite{Peotta_Torma}. This contribution originated from the topologically nontrivial geometric properties of Bloch states in momentum space and became particularly relevant in the regime where the pairing amplitude exceeded the quasi-flat band's energy width and the chemical potential \cite{Kopnin_Sonin_PhysRevLett, Peotta_Torma}. 
Here, we focus on the opposite regime of a small gap in the topologically trivial case, where the interplay between band dispersion, hybridization effects, and electronic interactions that drive superconductivity in the quasi-flat band remains, to the best of our knowledge, far from being understood.

In this work, we analyze superconductivity in a two-dimensional system described by a two-band heavy-fermion model, where a dispersive band hybridizes with a flat band, \cite{Song_PhysRevLett, Firoz_PhysRevB, Molecular_MATBG}. 
The resulting quasi-flat band features distinct momentum-dependent behavior: a nearly flat region at large momenta and a dispersive region near the Fermi momentum, as illustrated in Fig. (\ref{fig1}).
Two length scales play critical roles: short-range correlations, governing Cooper pair formation, and long-range correlations, essential for establishing phase stiffness.

We examine superconductivity within the framework of preformed Cooper pairs, \cite{Ohashi_progress, Strinati_physrep, Nature_Sacepe_Feigelman}. We calculate both the crossover temperature for pair formation and the BKT transition temperature, incorporating the effects of band hybridization. When the pairing amplitude exceeds the flat-band energy width, superconductivity collapses into a phase-incoherent state dominated by random phase fluctuations. In the opposite case, the BKT  temperature is governed by the Fermi surface contribution of the quasi-flat band and exhibits a dome-like dependence on doping.

\section{Two-band model}
We consider a minimal model of a hybridized flat band and conduction band system in two-dimensions, described by the Hamiltonian $\mathcal{H} =\sum_{s} \int  d{\bm r}  \Psi_{{\bm r},s}^{\dag} H_0({\bm r} )\Psi_{{\bm r},s} $, with
\begin{equation}\label{Model_H}
H_0({\bm r} ) = \left(-\frac{\boldsymbol{\partial}_r^2}{2m}+ \eta \right)  \frac{1+\sigma_z}{2} + g \sigma_x.
\end{equation}
Here $m$ is the effective electron mass in the conduction band, $g>0 $ represents the inter-band hybridization, and $\eta \gg g$ is the bandgap between the conduction band and the flat band. The Pauli matrices $\sigma_{x,z}$ describe the band degrees of freedom, and natural units $\hbar= k_{\mathrm{B}}=1$ are used throughout the paper. 
The hybridization parameter $g$ is treated as constant in both space and time.

\begin{figure}[t]
\centering
\includegraphics[width=8cm]{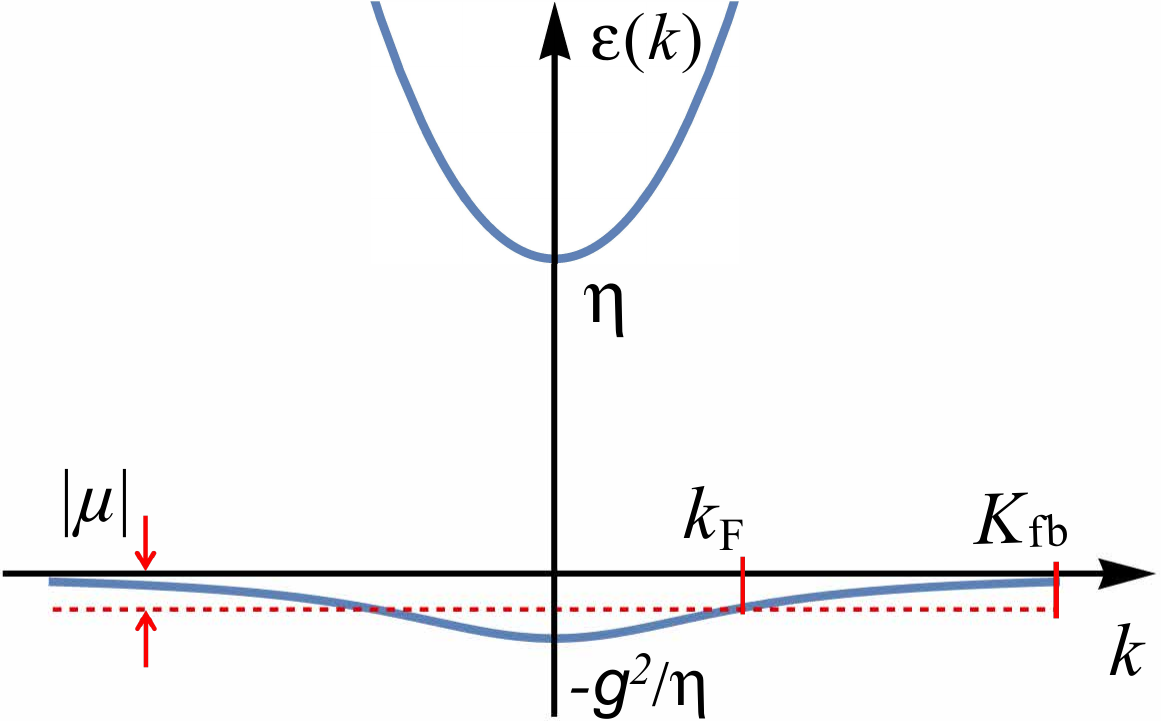}
\caption{\label{fig1} Illustration of the band dispersion $\epsilon({\bm k})$ for the upper parabolic band and lower quasi-flat band. At $g=0$, the lower band is entirely flat and is separated from the upper band by a gap $\eta$. 
For $g>0$, hybridization induces a nontrivial dispersion in the quasi-flat band. The bandgap is larger than the quasi-flat band's energy width, $\eta \gg g$. The momentum-space width of the quasi-flat band is denoted by $K_{\mathrm{fb}}$. The chemical potential $\mu<0$ determines the Fermi momentum $k_{\mathrm{F}}<K_{\mathrm{fb}}$.}
\end{figure}

The Fermi field operator is defined as
$\Psi_{{\bm r},s}  = (\Psi^{(c)}_{{\bm r},s},\Psi^{(f)}_{{\bm r},s})^{\mathrm{T}}$, where indices $(c)$ and $(f)$ denote conduction and flat-band states, respectively, and $s= \uparrow, \downarrow$ labels two spin projections. Diagonalization of  (\ref{Model_H}) yields the band dispersion
\begin{equation}
\epsilon_{\pm}({\bm k}) = \frac{1}{2}\left(\frac{{\bm k}^2}{2m}+\eta \right) \pm \sqrt{ \frac{1}{4}\left(\frac{{\bm k}^2}{2m}+\eta \right)^2 + g^2},
\end{equation}
also shown in Fig. (\ref{fig1}), where ${\bm k} = (k_x, k_y)$ is the momentum.

At $g=0$, the spectrum consists of a quadratic band given by $k^2/(2m)+\eta$ and a flat band at zero energy. For $g\neq 0$, a flat band develops a finite dispersion, which, in the limit $\eta \gg g$, can be approximated by
 \begin{equation}\label{Main_dispersion}
 \epsilon({\bm k}) \simeq -\frac{g^2}{\frac{{\bm k}^2}{2m}+\eta}. 
 \end{equation} 
Let us continue labeling this lower band as the quasi-flat band, as it has a weak finite dispersion through hybridization. 
 %%%%%%%%%%%%%%%%%
%%%%%%%%%%%%%%%%%
If the chemical potential $\mu$ is positioned within the quasi-flat band, so that the condition
\begin{equation}\label{Conditions_on_mu}
-\frac{g^2}{\frac{K_{\mathrm{fb}}^2}{2m}+\eta}>\mu> -\frac{g^2}{\eta}
\end{equation} 
is satisfied, we can define the Fermi momentum
$
k_{\mathrm{F}} = \sqrt{2m\left(g^2/|\mu| - \eta\right)}
$. 
Here, $k_{\mathrm{F}}$ is restricted to values smaller than the quasi-flat band's momentum width, $K_{\mathrm{fb}}> k_{\mathrm{F}}$.
Expanding the dispersion in (\ref{Main_dispersion}) near $k_{\mathrm{F}} $ yields
\begin{equation}\label{Spectrum_expansion}
\epsilon({\bm k}) -\mu \simeq  \frac{{\bm k}^2-k^2_{\mathrm{F}}}{2 m^*},
\end{equation}
where $m^*= m(g/\mu)^2$ is the effective mass of particles in the quasi-flat band. It is enhanced relative to that in the upper parabolic band by a factor of $(g/\mu)^2\gg 1$ increasing with $\mu \rightarrow 0$. 
The Fermi velocity $v_{\mathrm{F}}=  k_{\mathrm{F}}/m^*$ also varies as a function of doping reaching maximum at $\mu = -3g^2/(4\eta)$.
It is worth noting that the particle density of the quasi-flat band is given by the usual expression $N_{\mathrm{fb}} = k_{\mathrm{F}}^2/(2\pi)$. 
For $\mu \geqslant 0$, the quasi-flat band is fully occupied, yielding  $N_{\mathrm{fb}} = K_{\mathrm{fb}}^2/2\pi$. Otherwise, for $\mu< -g^2/\eta$, the band is empty, and $N_{\mathrm{fb}}=0$. We also recall that 
the dispersive band is empty for $\mu\lesssim \eta$.

We now consider the s-wave short-range pairing between electrons in the flat band, described by the interaction Hamiltonian $\mathcal{H}_{\mathrm{int}} = -\lambda \int d{\bm r} \Psi^{(f)\dag}_{{\bm r},\uparrow} \Psi^{(f)\dag}_{{\bm r},\downarrow}\Psi^{(f)}_{{\bm r},\downarrow}\Psi^{(f)}_{{\bm r},\uparrow}$, where $\lambda>0$ is the interaction constant. 
We neglect pairing in the upper dispersive band as well as the inter-band pairing. This approximation is based on the assumption that the chemical potential lies within the quasi-flat band and that the maximum amplitude of the attractive interaction potential $\propto \lambda K_{\mathrm{fb}}^2$ is smaller than the interband gap $\propto \eta$.
We also assume a generic attractive interaction mechanism of Cooper pairing without specifying its microscopic origin. 
The range of the pairing potential is considered to be smaller than the electron Fermi wave-length and the length-scale associated with the flat-band, $a <K^{-1}_{\mathrm{fb}}< k^{-1}_{\mathrm{F}}$. 
The discussion below can be readily extended to the situation when $K^{-1}_{\mathrm{fb}}< a<k^{-1}_{\mathrm{F}}$ by substituting the ultraviolet cut-off, $K^{-1}_{\mathrm{fb}}$, in the self-consistency equation in what follows with the scale $\propto a$.

Further analysis can be performed by introducing the complex field $\Delta({\bm r})=-\lambda \langle \Psi^{(f)}_{{\bm r}, \downarrow} \Psi^{(f)}_{{\bm r}, \uparrow}  \rangle$ via the Hubbard - Stratonovich decoupling of the interaction term. The Bogolyubov - de Gennes Hamiltonian is given by
\begin{eqnarray}\label{BdG_Hamiltonian_main}
H({\bm r}) = 
 \left( \begin{matrix}
H_0({\bm r}) - \mu & - \Delta({\bm r})(1- \sigma_z)/2  \\
- \Delta({\bm r})^*(1- \sigma_z)/2  & -H_0({\bm r}) + \mu
 \end{matrix}\right)
\end{eqnarray}
acting on the vector of Nambu field operators $\Psi_{{\bm r}}  = (\Psi^{(c)}_{{\bm r},\uparrow},\Psi^{(f)}_{{\bm r},\uparrow}, -\Psi^{(c)\dag}_{{\bm r},\downarrow}, -\Psi^{(f)\dag}_{{\bm r},\downarrow})^{\mathrm{T}}$.
At $g/\eta \ll 1$, however it suffices to project the Hamiltonian to the low-energy quasi-flat band $H\rightarrow H^{(f)}$, where
\begin{eqnarray}\label{BdG_Hamiltonian_f}
H^{(f)}({\bm r}) = 
 \left( \begin{matrix}
\epsilon({\bm r}) - \mu & - \Delta({\bm r})  \\
- \Delta^*({\bm r})  & -\epsilon({\bm r}) + \mu
 \end{matrix}\right)
\end{eqnarray}
is a $2\times 2$ Hamiltonian. We neglect small gradient-dependent corrections to the gap function compared to the regular quasi-flat band dispersion contribution, taking $\Delta (1- g^2/\eta^2) \approx \Delta$ (the eigenvalues of (\ref{BdG_Hamiltonian_main}) are given in the Appendix).
Diagonalizing $H^{(f)}$, one obtains the following dispersion
\begin{equation}\label{Spectrum_hybridization}
E({\bm k}) = \sqrt{(\epsilon({\bm k})  - \mu)^2 + |\Delta|^2},
\end{equation}
with the large-$k$ asymptotic given by
$
E({\bm k})|_{k \rightarrow \infty} = \sqrt{\mu^2+|\Delta|^2}
$.

To proceed, within the preformed Cooper pair model  \cite{Nature_Sacepe_Feigelman}, we first investigate the crossover into the inhomogeneous Cooper pair state and then analize the transition to a globally coherent superconducting state calculating the BKT temperature.

\section{Isolated flat band}
It is instructive to begin with a simplified model, where we assume zero hybridization between the bands by setting $g=0$.
In this situation, the transport and thus superconductivity are absent due to strong localization of particles within the flat band. However, there may be a tendency to local Cooper pairing, with each bound pair's phase being uncorrelated across different regions. This scenario aligns with the model of superconducting grains in the limiting situation with no inter-grain correlations, \cite{Efetov_granular}. Leaving the analysis of time-dependent fluctuations of the pairing function to Appendix (\ref{sec:Appendix_B}), here we focus on the static limit.

We can divide the system into an array of domains of uniform size, with area determined by the inverse of flat-band width in momentum space 
\begin{equation}
\int_{k<K_{\mathrm{fb}}} \frac{d{\bm k}}{(2\pi)^2} = \frac{K^2_{\mathrm{fb}}}{4\pi}.
\end{equation}
The pairing function $\Delta({\bm r})$ may be coarse-grained over each of these domains as
$
\Delta({\bm r})\rightarrow \Delta_j = |\Delta_j| e^{i\theta_j}.
$
The local pairing $\Delta_j$ in each domain $j$ is thus uniform within each domain, but its phase $\theta_j$ (and potentially the amplitude $|\Delta_j|$) can vary between domains. 
However, for simplicity, we assume that variations in $|\Delta_j|\equiv |\Delta|$ across  domains are negligible, assuming the absence of spatial disorder in the attractive interaction.

Neglecting the electromagnetic term, the action for an isolated flat band in the static approximation at temperature $T$ is given by $S= \sum_j S_j$, where summation is over all domains and
\begin{equation}\label{toy_action}
T S_j= \frac{4\pi |\Delta|^2}{\lambda K^2_{\mathrm{fb}}} + \mu - 2T\ln\left(\cosh \frac{ \sqrt{\mu^2+|\Delta|^2}}{2T}\right).
\end{equation}
At low temperatures, this simplifies to
$T S_j = \frac{4\pi |\Delta|^2}{\lambda K^2_{\mathrm{fb}}} -  \sqrt{\mu^2+|\Delta|^2} + \mu$. 
The convergence of $S$ at large $|\Delta|$ is ensured by the $|\Delta|^2$ term, unlike in traditional Ginzburg-Landau theory, where the quartic term $\sim |\Delta|^4$ provides stability.
Furthermore, here the condensation energy at charge neutrality is given by $E_{\mathrm{cond}} = - |\Delta|/2$.

The self-consistency equation for $\Delta_j = |\Delta| e^{i\theta_j}$ can be derived from (\ref{toy_action}) as          
\begin{equation}\label{Trivial_SC}
|\Delta| = \frac{\lambda K^2_{\mathrm{fb}}}{8\pi} \frac{|\Delta|}{ \sqrt{\mu^2+|\Delta|^2}} \tanh\left( \frac{ \sqrt{\mu^2+|\Delta|^2}}{2T}\right).
\end{equation}
We note that Eq. (\ref{Trivial_SC}) defines $\Delta_j$ up to an arbitrary phase $\theta_j$ in each domain. There is no global phase coherence established across the system.

For an isolated flat band, it is convenient to re-express the chemical potential in terms of the flat-band filling factor by using the expression for the particle density, which is here given by
$
N_{\mathrm{fb}} = \frac{K^2_{\mathrm{fb}}}{4\pi}\left[ 1+ (\mu/ \sqrt{\mu^2+|\Delta|^2}) \tanh ( \sqrt{\mu^2+|\Delta|^2}/2T)\right].
$
This allows one to introduce the filling factor $0\leq \nu\leq 1$ as $\nu = 2\pi N_{\mathrm{fb}}/ K^2_{\mathrm{fb}} $.
Combining Eq. (\ref{Trivial_SC}) with the equation for the number of particles, one rewrites the chemical potential at finite $|\Delta|$ as 
\begin{equation}\label{mu_main}
\mu = 2T_{\mathrm{PG}} (2\nu -1),
\end{equation}
where $T_{\mathrm{PG}} = \lambda K^2_{\mathrm{fb}}/16\pi$.
As a result, the gap amplitude is given by
\begin{align}\label{Cases_2}\nonumber
&|\Delta| \propto T_{\mathrm{PG}}(\nu) \sqrt{(1-\nu )\nu} \sqrt{1-\frac{T}{T_{\mathrm{PG}}(\nu) }},~ T\rightarrow T_{\mathrm{PG}}(\nu)\\
&|\Delta| = 4T_{\mathrm{PG}}\sqrt{(1-\nu )\nu},~~~~~~~~~~~~~~~~ T=0,
\end{align}
where
$
T_{\mathrm{PG}}(\nu)  =  \frac{2\nu -1}{\mathrm{arcth}(2\nu -1)} T_{\mathrm{PG}}
$ is the filling-factor dependent crossover temperature.
This means that $T_{\mathrm{PG}}$ is the crossover temperature at the half-filling of the flat-band. The solution is similar to the one first obtained for the flat-band superconductivity by Khodel and Shaginyan 
\cite{Khodel_Shaginyan, Nozieres} and recently revisited in \cite{Peotta_Torma}. 

Let us emphasize once more that both $T_{\mathrm{PG}}(\nu)$ and the pairing amplitude $|\Delta|$ are directly proportional to the interaction constant $\propto \lambda$ and reach a maximum at half-filling \cite{Khodel_Shaginyan}. The latter is also similar to the theoretical result for superconductivity at very strong magnetic fields \cite{MacDonald_quantumhall}. 

After this extended discussion, we now turn our attention to superconductivity, taking into account hybridization between the bands. 

\section{Hybridization between the bands}
\subsection{Self-consistency equation}
In the two-dimensional pairing problem, the many body ground state is unstable to s-wave pairing if and only if a two-body bound state exists,  \cite{Bashkin_tc, Miyake_tc, PhysRevLett_Randeria}. In our case, we have two hybridized bands, so we expect three two-body bound states: two for particles from the same bands (with bound energies positioned at approximately $2\eta$ and $-2g/\eta$) and one for particles from different bands, positioned at $\approx \eta$. The detailed derivation of the bound state energies is given in Appendix. Let us here model the pairing potential as $U({\bm r})=-U_0 \theta(a-r)$, where $U_0>0$ and $a$ is the radius of the pairing potential, such that $\lambda \propto a^2 U_0$. The bound state energy of two particles from the quasi-flat band, assuming $2 g^2/\eta >  U_0$, is given by $\epsilon_3 = - \frac{2g^2}{m a^2\eta^2}\exp\left( - \frac{2}{m a^2 U_0}\frac{g^2}{\eta^2} \right) $, measured from the bottom of the quasi-flat band at $ -2g^2/\eta$. A bound state exists for an arbitrarily weak potential. 

We note that the two-body problem in a three-dimensional superconducting system described by two parabolic bands in the strongly interacting case was recently studied numerically in Ref. \cite{Perali_3}. 
In particular, a two-particle bound-state solution which has components in both bands was found in the presence of the inter-band interaction. 
Here, we focus on the formation of the two-body bound state in the quasi-flat band and in the two-dimensional system.

The many-body ground state may be qualitatively analyzed with the self-consistency equation for the pairing amplitude. 
Incorporating hybridization between the bands, we apply a continuous approximation, converting the summation over the grains into an integral, and obtain
\begin{align}\label{SC_equation_2}
\frac{|\Delta|}{\lambda} = \int_{k<K_{\mathrm{fb}}}  \frac{d{\bm k}}{(2\pi)^2} \frac{|\Delta|}{2E({\bm k})} \tanh\left( \frac{E({\bm k})}{2T}\right),
\end{align}
where it now contains a momentum dependent band dispersion $E({\bm k})$ defined by Eq. (\ref{Spectrum_hybridization}).
The self-consistency equation shall be supplemented with the equation for the number of particles
\begin{eqnarray}\label{number_eq}
N_{\mathrm{fb}} = \int_{k<K_{\mathrm{fb}}} \frac{d{\bm k}}{(2\pi)^2}\left\{1- \frac{\epsilon({\bm k})-\mu}{E({\bm k})}\tanh\left(\frac{E({\bm k})}{2T}\right)\right\}~~~
\end{eqnarray}
in order to self-consistently determine the chemical potential.

Eqs. (\ref{SC_equation_2}) and (\ref{number_eq}) can be solved perturbatively in both strong and weak interaction regimes.
We first seek for a solution of  Eq. (\ref{SC_equation_2}) in the linearized limit and when the temperature is smaller than the energy width of the quasi-flat band,  $g^2/\eta - |\mu|> T$ and $|\mu|> T$. 
In this case, Eq. (\ref{number_eq}) gives a small temperature dependent corrections to the chemical potential so that we can proceed with the self-consistency equation only. We find
\begin{eqnarray}\label{Sc_second}
\frac{1}{\lambda} = \frac{1}{\lambda_c}+
\frac{m^*}{2\pi}
\ln\left| \frac{\gamma}{\pi} \frac{ k_{\mathrm{F}}}{m^* T}  \sqrt{K^2_{\mathrm{fb}}-k_{\mathrm{F}}^2} \right|,
\end{eqnarray}
with $\lambda^{-1}_c = \frac{K^2_{\mathrm{fb}} - 2 k^2_{\mathrm{F}}}{8\pi |\mu|} $ and $\gamma = e^{C}\simeq 1.78$, where $C \simeq 0.577$ is the Euler constant. 

The first term on the right hand side of (\ref{Sc_second}) exhibits an ultraviolet divergence arising from the large-momentum region of quasi-flat band. The power-law divergent contribution is analogous to that obtained in three dimensional systems. For a review on the latter, see Refs. \cite{Ohashi_progress, Strinati_physrep}. However, the logarithmic dependence on the momentum cutoff in the second term in (\ref{Sc_second}) is a characteristic of two-dimensional systems with parabolic dispersion, such as in the problem of two-particle bound state formation (or dimer formation in the context of thin film $\mathrm{^3He}$)  \cite{Bashkin_tc, Miyake_tc, PhysRevLett_Randeria}.

\begin{figure}[t]
\centering
\includegraphics[width=6cm]{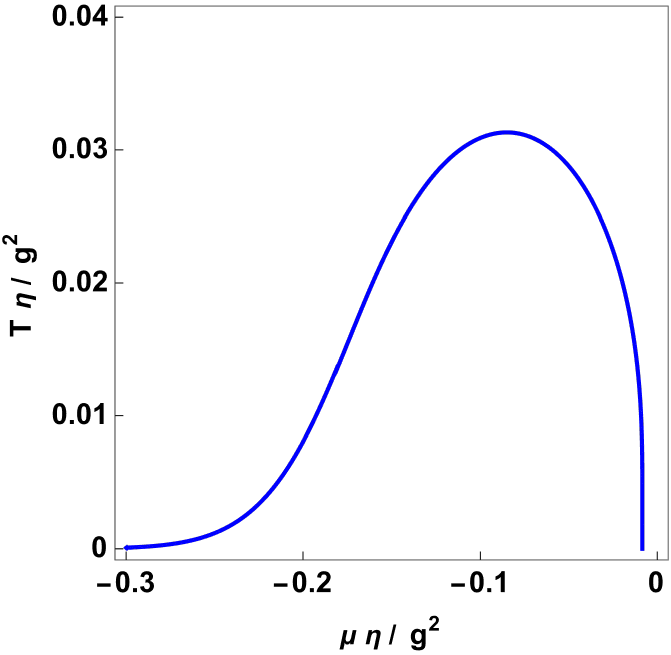}
\includegraphics[width=6cm]{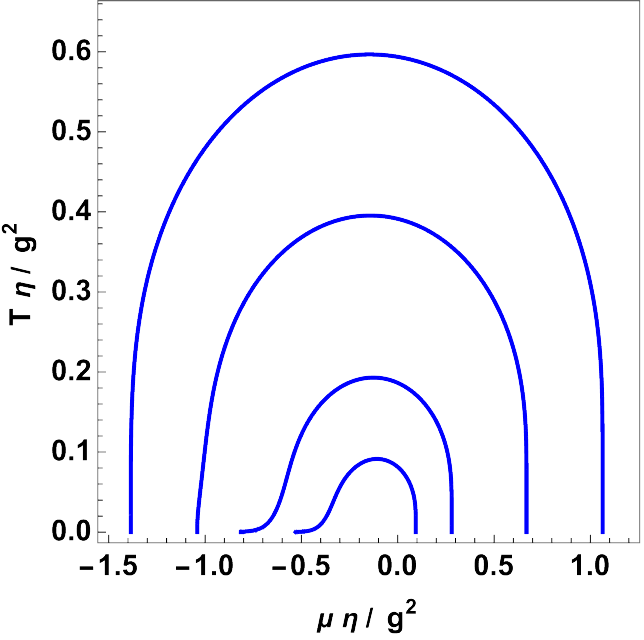}
\caption{\label{fig2} The solution of the self-consistency Eq. (\ref{SC_equation_2}) for the band momentum-width $K^2_{\mathrm{fb}}/(2m\eta) = 20$ and interaction parameter $T_{\mathrm{PG}}\eta/g^2 = 0.04$ (top) and $T_{\mathrm{PG}}\eta/g^2 = (0.1,0.2,0.4,0.6)$ (bottom). The chemical potential and the temperature are normalized to the energy width of the quasi-flat band, $g^2/\eta$. 
Bottom: The system transitions into the strong interaction regime.}
\end{figure}

The weak interaction regime  is defined by the range of parameters, that satisfy $\lambda< \lambda_c$, see Fig. (\ref{fig2}).
The critical temperature is given by the BCS expression
\begin{eqnarray}\label{T_bcs}
T_{\mathrm{BCS}} = \frac{\gamma}{\pi} \frac{k_{\mathrm{F}} }{m^*}  \sqrt{K^2_{\mathrm{fb}}-k_{\mathrm{F}}^2} \exp\left\{-\frac{2\pi}{\lambda m^*}\left(1-\frac{\lambda}{\lambda_c} \right) \right\}.~~~~
\end{eqnarray}
The prefactor in (\ref{T_bcs}) vanishes both at the top ($\mu \rightarrow 0$) and bottom ($\mu \rightarrow -g^2/\eta$) of the quasi-flat band,  
compared with the chemical potential dependence in the Gorkov and Melik-Barkhudarov result for critical temperature for an unbounded parabolic band \cite{Barkhudarov_Tc}. We emphasize that here $g^2/\eta > |\mu|$ so that when $g\rightarrow 0$, we go back to the results outlined in the previous section.
Furthermore, the exponent contains the large effective mass scaling as $m (g/\mu)^2$, along with the renormalized interaction constant.
Finally, the pairing amplitude at zero temperature can be shown to follow the standard relationship with the critical temperature 
$|\Delta| = \pi T_{\mathrm{BCS}}/\gamma$.

In the strong interaction case ($\lambda >\lambda_c$), the logarithmic temperature dependence of the BCS term in Eq. (\ref{Sc_second}) fails to provide a consistent solution.
Hence, the temperature-dependence of the leading $\lambda_c$-term together with the number equation (\ref{number_eq}) must be reconsidered. 

%%%%%%%%%%%%%%%%%
%%%%%%%%%%%%%%%%%

We note that the crossover between the BCS state and Bose state of bound pairs in the two dimensional systems with parabolic dispersion was analyzed in Refs. \cite{Bashkin_tc, Miyake_tc, PhysRevLett_Randeria}. 
Here we have an additional large-momentum flat-band contribution to the self-consistency and particle number equations. Depending on the parameters of the band structure and doping there is a conversion between two limiting cases of strong interaction regime in which either of two contributions dominate. However, the expression for the gap amplitude is not as simple as in the parabolic band case.  

For example under the condition $K_{\mathrm{fb}}^2/(2m) >g^2/\eta$,
the large-momentum contribution determine the formation of bound pairs (or dimers). We recover Eq. (\ref{Trivial_SC}) with a small hybridization induced correction. Specifically at $\mu=0$, we obtain  
\begin{equation}
T_{\mathrm{PG}} = \frac{\lambda K^2_{\mathrm{fb}} }{16\pi} \left(1- \frac{4 m g^2}{\eta K^2_{\mathrm{fb}}} \right). 
\end{equation}
Here the temperature of bound state formation is again linearly proportional to the interaction constant. To proceed, for the superconducting state, we must further identify the BKT temperature.

\subsection{Phase stiffness and BKT temperature}
From the extended BdG Hamiltonian (\ref{BdG_Hamiltonian_main}), we determine the general expression for the current operator (no projection to the flat-band)
$
\hat{{\bm J}}({\bm r}) = (e/m) \left(-i\boldsymbol{\partial}_r  - e{\bm A}({\bm r}) \tau_z \right)(1+\sigma_z)/2,
$
where ${\bm A}({\bm r})$ is the vector potential, $e<0$ is the electron charge, and $\tau_{z}$ is the Pauli matrix acting in Nambu space.
Using the textbook definition \cite{AGD}, the current density can be written in components as $J_{\alpha}({\bm q}) = - (2e / \hbar)^2Q_{\alpha \beta}({\bm q})  A_{\beta}({\bm q})$ (here $\hbar$ is restored for dimensionality).
The kernel $Q_{\alpha\beta}({\bm q}) $ is determined by the hybridization induced dispersion of the quasi-flat band and vanishes at $g=0$.
We consider the London limit and take $Q_{\alpha\beta}(0) = Q \delta_{\alpha\beta}$ at zero wave-vector. Leaving the derivation of $Q$ to Appendix (\ref{sec:Appendix_A}), let us examine two specific limiting cases.

\subsubsection{Large pairing amplitude $\eta\gg |\Delta|> g^2/\eta$}
\begin{figure}[t]
\centering
\includegraphics[width=7cm]{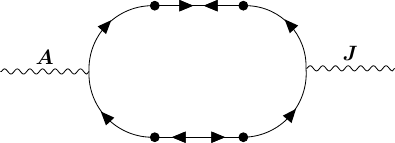}
\caption{\label{fig3} A typical diagram illustrating the lowest-order inter-band hybridization contribution in the large pairing amplitude limit to the supercurrent ${\bm J}$ as a response to the vector potential ${\bm A}$ in (\ref{Q_strange}). Dots represent the hybridization $\propto g$ between the non-superconductive dispersive upper band and the lower flat band with Copper pairing. Single-arrow lines denote the normal Green function in the upper band, while the double-arrow lines indicate the anomalous Green functions in the flat-band.}
\end{figure}

Consider a toy-model situation where $|\Delta|$ is larger than the energy width of the quasi-flat band. 
Focusing on contributions in the lowest order in $g$, we find the following expression for the phase stiffness
\begin{eqnarray}\nonumber\label{Q_strange}
Q &=& \frac{g^4 |\Delta|^2}{2m^2}T\sum_n\frac{1}{(\omega_n^2+ |\Delta|^2+ \mu^2)^2}
\\ 
&\times& \int \frac{d{\bm k}}{(2\pi)^2} \frac{k^2}{\left[\omega_n^2+\left(\frac{k^2}{2m}+\eta-\mu\right)^2\right]^2}.
\end{eqnarray}
To illustrate, the Feynman diagram describing the lowest order contribution to supercurrent is shown in Fig. (\ref{fig3}). 
Next, we approximate the integral over momenta noting $\eta \gg |\Delta|,T$ which yields
$Q = \frac{g^4 |\Delta|^2}{48\pi \eta^2 E_0^3}\frac{\sinh(E_0/T)-E_0/T}{\cosh^2(E_0/2T)}$.
Taking the zero temperature limit
and using the expressions in (\ref{mu_main}) and (\ref{Cases_2}) for the self-consistent chemical potential and gap amplitude, we find
$
Q  = \frac{(1-\nu)\nu g^4}{24\pi \eta^2 T_{\mathrm{PG}}}
$.
The phase stiffness is inversely proportional to the interaction constant and exhibits a dome-like dependence on doping, with a maximum at half-filling. A similar result was shown numerically for superconductivity in a model with two overlapping parabolic bands with light and heavy effective electron masses in Ref. \cite{Perali_1}. However, this contrasts with the quantum metric contribution to the phase stiffness in two-dimensional systems with Dirac dispersion at low doping, where $Q\sim | \Delta|$ at $|\Delta| \gg |\mu|$, \cite{Kopnin_Sonin_PhysRevLett, Guodong_arxiv}. 

Applying the definition of the BKT temperature $T_{\mathrm{BKT}} = (\pi/2) Q|_{T=T_{\mathrm{BKT}}}$,  \cite{B, KT}, we obtain 
\begin{equation}\label{TBKT_verystrong}
T_{\mathrm{BKT}} =   \frac{(1-\nu)\nu  g^4}{48 \eta^2T_{\mathrm{PG}}}.
\end{equation}
The BKT temperature is much smaller than the temperature of Cooper pair formation $T_{\mathrm{BKT}} \ll T_{\mathrm{PG}}, |\Delta|$, which justifies the zero temperature limit taken for $Q$ here. 
Importantly, although both the temperature $T_{\mathrm{PG}}$ and the pairing amplitude $|\Delta|$ increase with stronger interactions, $T_{\mathrm{BKT}}$ decreases as the interaction strength increases.
In other words superconducting state shrinks while inhomogeneous states expands as the interaction strength increases. 

%%%%%%%%%%%%%%%%%%%%%% BEGIN FIGURE %%%%%%%%%%%%%%%%%%%
%%%%%%%%%%%%%%%%%%%%% END FIGURE %%%%%%%%%%%%%%%%%%%

\subsubsection{Small pairing amplitude $g^2/\eta \gg |\Delta|$}
In a more realistic scenario, the energy width of the quasi-flat band exceeds the pairing amplitude and the temperature, such that $g^2/|\eta|>|\mu| > |\Delta|,T$. 
Performing projection to the quasi-flat band, we obtain superconducting phase stiffness in the usual form \cite{AGD}:
\begin{eqnarray}\label{Q_main_2}
Q = \frac{k_{\mathrm{F}}^2}{8 m^*} T\sum_n  \frac{|\Delta|^2}{(\omega_n^2+|\Delta|^2)^{3/2}}.
\end{eqnarray} 
However, we recall that $|\Delta|$ follows a dome-like dependence on doping, as described by (\ref{Cases_2}) and (\ref{T_bcs}) for the strong and weak interaction cases, respectively, which indirectly influence the doping and temperature dependencies of $Q$. 
Specifically, at $T\ll |\Delta|$, (\ref{Q_main_2}) simplifies to
$
Q = k_{\mathrm{F}}^2/(8\pi m^*)
$, so that the BKT temperature in this case is given by
\begin{equation}\label{notstrong_BKT}
T_{\mathrm{BKT}} = \frac{k_{\mathrm{F}}^2}{16 m^*}.
\end{equation}  
There is an expected dependence on the effective electron mass $m^* = m (g/\mu)^2$ and concentration of particles $k_{\mathrm{F}}^2/(2\pi)$, which however shows a dome-like behavior as a function of doping, with a maximum value 
$
T^{*}_{\mathrm{BKT}} = g^2/(32\eta) 
$
at $\mu = -g^2/(2\eta)$.
The chemical potential that maximizes the BKT transition temperature  generally differs from the one that maximizes the pair formation temperature.
To further enhance the $T^{*}_{\mathrm{BKT}}$, it is necessary to broaden the energy bandwidth of the quasi-flat band.
We note that this maximum value is only numerically smaller than $|\mu|$, which aligns with the results obtained for superconductivity in the strong coupling regime in graphene layers \cite{Kopnin_Sonin_PhysRevLett} and in metals with parabolic band dispersion \cite{Zyuzin_parabolic}. 

Let us compare the BKT temperature with the gap amplitude in the weak interaction regime
\begin{equation}
\frac{T_{\mathrm{BKT}}}{|\Delta|} = \frac{1 }{16}  \frac{K_{\mathrm{fb}}}{k_{\mathrm{F}}} \exp\left\{-\frac{2\pi}{\lambda m^*}\left(1-\frac{\lambda}{\lambda_c} \right) \right\}.
\end{equation}
For $|\Delta|>T_{\mathrm{BKT}}$, the system is in the regime of strong phase fluctuations consistent with the pseudogap of preformed pairs.

To conclude this section, we want to point out that the phase stiffness of a fully gapped superconductor at $\eta \gg |\Delta|$ exhibits an exponentially weak temperature dependence at low temperatures. 
However, the influence of band anisotropy, which is neglected here, can suppress the pairing amplitude along specific momentum-space directions. This anisotropy enhances dissipation channels that impact phase fluctuations in the dissipative regime. Consequently, it may result in a power-law temperature dependence of the phase stiffness at low temperatures, as described by the two-fluid model of superconductivity in Ref. \cite{PhysRevB_Benfatto}.

Furthermore, the $T_{\mathrm{BKT}}$ is sensitive to disorder in the system, neglected in our discussion, being strongly suppressed by electron scattering by impurities compared to $T_{\mathrm{BCS}}$. 
For example \cite{AGD}, in the weak interaction regime, $Q = \frac{k_{\mathrm{F}}^2}{8 m^*} \tau |\Delta| \tanh(|\Delta|/2T)$, where $\tau$ is the mean free time due to electron scattering by disorder in the quasi-flat band. A detailed analysis of phase fluctuations in the superconducting system can be found in Ref. \cite{Larkin_book}.  We also note that the suppression of thermal superconducting fluctuations in a quasi-one-dimensional band due to inter-band interaction with a higher-dimensional band in the multi-band case was recently studied in Ref. \cite{Perali_2}.

\section{Discussion}
Let us now estimate the BKT temperature for the case when the pairing amplitude is smaller than the energy width of the quasi-flat band $g^2/\eta> |\Delta|, T_{\mathrm{PG}}$, considering parameters relevant for twisted bilayer graphene (TBG) as a particular example. 
TBG might be considered within the heavy-fermion model proposed in Ref. \cite{Song_PhysRevLett}. In that model, the dispersion around the Dirac cones ($K$ and $K’$ points of the moire Brillouin zone) is assumed to be completely flat, while the dispersion around the $\Gamma$-point is parabolic. This parabolic dispersion originates from the hybridization between the flat and conducting bands. Within the preformed pairing scenario, the Cooper pairs in TBG are concentrated near AA-stacking regions in the TBG moire pattern overlapping over the intermediate AB-regions, \cite{Heikkila_twisted, Wu_Macdonald_twisted, Song_PhysRevLett, Firoz_PhysRevB, Molecular_MATBG}.  We assume that the conduction and valence subbands of the quasi-flat band in TBG are gapped at the charge neutrality, so that these bands may be considered separately.

Since the BKT temperature was experimentally observed to be much smaller than the chemical potential in TBG, we assume that the dominant contribution to the phase stiffness comes from the band dispersion around the $\Gamma$-point of TBG moire Brillouin zone \cite{Trambly_MATBG, PNAS_macdonald}. We suggest that superconductivity in TBG is governed by phase fluctuations so that the expression for the crossover temperature (\ref{T_bcs}) at $\lambda \rightarrow \lambda_c$ and BKT temperature (\ref{notstrong_BKT}) might be qualitatively applied. 

As a remark, we note that in the extreme limit, $K_{\mathrm{fb}}$ may extend to the edges of the Brillouin zone, corresponding to the lattice cites of the hopping model. However, in general, $K_{\mathrm{fb}}$ can represent much wider regions in real space. For example, it may correspond to the width of the AA regions in the twisted bilayer graphene, which serve as an effective lattice cites in the moire lattice.

To estimate the width of the quasi-flat band in momentum space, $K_{\mathrm{fb}}$, we may utilize the parameters of the moire Brillouin zone in TBG. 
Taking $K_{\mathrm{fb}}$ to be the distance between $\mathrm{K}$ and $\Gamma $ points of moire Brillouin zone, we obtain $K_{\mathrm{fb}}\sim  \frac{4\pi}{3b}\theta \simeq 0.3 \mathrm{nm}^{-1}$ for $b=0.25 \mathrm{nm}$ and $\theta=1^{\circ}$. Using parameters and notations from Ref. \cite{Song_PhysRevLett}, we take the energy separation between dispersive and quasi-flat bands to be proportional to the bands separation at the $\Gamma$-point: $\eta \equiv |\gamma | \simeq 25 \mathrm{meV}$. Mapping the dispersions 
$ |\epsilon({\bm k})| = \frac{g^2/\eta}{1+{\bm k}^2/(2m\eta)} \sim \frac{M}{1+ v_{\star}^2{\bm k}^2/\gamma^2}$, where $M=4~\mathrm{meV}$ and $|v_{\star}| = 4 ~\mathrm{eV} \r{A}$, we find $1/\sqrt{2m\eta} \simeq 16~\mathrm{nm}$ in order to estimate $k_{\mathrm{F}} = \sqrt{2m\eta}\left[g^2/(|\mu|\eta) -1\right]^{1/2}$. Further, we set the chemical potential $|\mu|=2~\mathrm{meV}$ within the quasi-flat band with energy width $g^2/\eta = 4~ \mathrm{meV}$ (here $|\mu|\simeq 0.5\mathrm{meV}$ corresponds to half-filling) and obtain $K_{\mathrm{fb}}\sim 5k_{\mathrm{F}}$. So that the large-momentum and Fermi surface contributions are strongly mixed in the self-consistency Eq. (\ref{Sc_second}) for crossover temperature. On the other hand, the maximal BKT temperature can be estimated as $T^{*}_{\mathrm{BKT}} = g^2/(32 \eta) \simeq 1.5 \mathrm{K}$.

We can also estimate the magnitude of perpendicular magnetic field required for the vortex lattice melting at low temperatures.
The lower critical field is vanishingly small in thin films $ B_{c1} \sim \frac{\Phi_0}{8\pi \lambda^2} \frac{d}{\max{(R, \lambda)}}
$, where $\Phi_0$ is the flux quantum, $
\lambda = \sqrt{ \frac{c^2}{4\pi e^2}\frac{ dm^*}{ N_{\mathrm{fb}}}}
$ is the in-plane London penetration length, $d$ is the thickness of the thin film superconductor, and $R$ is the typical length of the film.  Taking $d \simeq 1 \mathrm{nm}$, we find $\lambda \simeq 2 \mathrm{\mu m}$.
To qualitatively estimate a second critical field $B_{c2}$, using band dispersion parameters from above we find the Fermi velocity $v_{\mathrm{F}} \simeq 5 \cdot 10^{6} ~\mathrm{cm/s}$, which is of the same order as the phonon velocity in an isolated graphene sheet. It gives coherence length $\xi = v_{\mathrm{F}} /(2\pi T_{\mathrm{BKT}}) \simeq 44 ~\mathrm{nm}$ resulting in $B_{c2} = \Phi_0/(2\pi \xi^2)\simeq 170~\mathrm{mT}$. 

The estimations are consistent with the upper critical field, superconducting transition temperature, and pseudogap temperature in TBG observed in transport and scanning tunnelling microscope experiments, \cite{Exp_TBG_superconductor, Yazdani_nature}. For a summary of the experimental results for the superconducting critical magnetic field and temperature, see Ref. \cite{CriticalField_Macdonald}.

Finally, we note that the superconducting stiffness may be sensitive to the Zeeman pair breaking effect of the magnetic field. The Hamiltonian (\ref{Model_H}) with an additional Zeeman contribution from the in-plane magnetic field $B$ is given by
\begin{eqnarray}
H_0+
 \left( \begin{matrix}
\frac{g_c}{2} \beta_B B\sigma_z & 0  \\
0  &\frac{g_{\mathrm{fb}}}{2} \beta_B B\sigma_z 
 \end{matrix}\right),
\end{eqnarray}
where $\beta_B$ is the Bohr magneton and $g_{c,\mathrm{fb}}$ are the electron g-factors in the dispersive and flat bands. Dispersion of the quasi-flat band then reads
$\epsilon_{{\bm k}} =  \frac{1}{2m^*} ({\bm k}^2-k^2_{\mathrm{F}}) \pm g^*  \beta_B B/2$, where $g^* = g_{\mathrm{fb}} +(\mu/g)^2 (g_{\mathrm{c}}-g_{\mathrm{fb}}) $. 
Generally, the g-factor of the quasi-flat band electrons is renormalized due to hybridization.
To incorporate the effect of the Zeeman term $|\mu| \gg |g^*\beta_B B| $,  we substitute $\omega_n \rightarrow \omega_n- i \left|g^*\right|\beta_B B/2$ in (\ref{Q_main_2}), which results in the usual expression for the phase stiffness in the  zero temperature limit
\begin{eqnarray}\label{Q_main_LOFF}
Q =\frac{k_{\mathrm{F}}^2}{8\pi m^*}  \left\{1- \mathrm{Re} \frac{\left|g^*\beta_B B/2\right|}{\sqrt{(g^*\beta_B B/2)^2-|\Delta|^2}} \right\}.
\end{eqnarray} 
The negative sign of the phase stiffness signals of the instability towards LOFF state. The complete phase diagram of the quasi-flat band superconductor in the magnetic field is an interesting separate research topic \cite{FFLO_BKT} and beyond the scope of the present paper.

To summarize, we studied superconductivity in the two-dimensional quasi-flat band system within the preformed Cooper pair model. We calculated the Cooper pairing crossover temperature and the BKT temperature as a function of doping and interaction strength.
If the gap amplitude is larger than the energy width of the quasi-flat band, the strong interaction regime, the superconducting state vanishes while the inhomogeneous state expands linearly as the interaction strength increases. 
However, in the opposite case, the BKT temperature exhibits a dome-like dependence on doping. 
\\
\emph{\textbf{Acknowledgements}}. 
We thank the Pirinem School of Theoretical Physics for warm hospitality. The numerical evaluations that support the findings in Fig. (\ref{fig2}) are available from the authors upon reasonable request.

%%%%%%%%%%%%%%%%%%%%%%%%%%%
\bibliography{Flatband_references_update.bib}
%%%%%%%%%%%%%%%%%%%%%%%%%%%
\clearpage
\onecolumngrid
\begin{center}
\rule{0.38\linewidth}{1pt}\\
\vspace{-0.37cm}\rule{0.49\linewidth}{1pt}
\end{center}
\setcounter{section}{0}
\setcounter{equation}{0}
\renewcommand{\theequation}{\Alph{subsection}.\arabic{equation}}

\section*{Appendix} 
\subsection{\label{sec:Appendix_A}Green functions and supercurrent}
With Hamiltonian (\ref{BdG_Hamiltonian_main}), the Green function in Nambu representation is given by 
$\mathcal{G}(i\omega_n,{\bm k}) = [i\omega_n - H({\bm k})]^{-1}$. In the general form
\begin{eqnarray}
\mathcal{G}(i\omega_n,{\bm k}) = 
 \left( \begin{matrix}
i\omega_n - \epsilon_1({\bm k}) & -g&0 &0 \\
-g  &i\omega_n -\epsilon_2 &0& \Delta\\
0&0  &i\omega_n +\epsilon_1({\bm k}) & g\\
0&\Delta^*  &g&i\omega_n +\epsilon_2 \\
 \end{matrix}\right)^{-1}.
\end{eqnarray}
In our model we take $$ \epsilon_1({\bm k}) = \frac{k^2}{2m} + \eta-\mu$$ and $$ \epsilon_2= -\mu.$$
At $\Delta =0$, the eigenvalues are given by
\begin{eqnarray}
\epsilon_{\pm}({\bm k}) = \frac{\epsilon_1({\bm k})+\epsilon_2}{2}\pm \sqrt{\frac{(\epsilon_1({\bm k})-\epsilon_2)^2}{4}+g^2}
\end{eqnarray}
In our notations
\begin{eqnarray}
\epsilon_{\pm}({\bm k}) = \frac{1}{2}\left(\frac{{\bm k}^2}{2m}+\eta \right) \pm \sqrt{ \frac{1}{4}\left(\frac{{\bm k}^2}{2m}+\eta \right)^2 + g^2} -\mu,
\end{eqnarray}
At finite $\Delta$, the eigenvalues are
\begin{eqnarray}
E^2_{\pm}({\bm k}) =\frac{1}{2} (|\Delta|^2+ \epsilon_1^2({\bm k})+\epsilon_2^2) + g^2 \pm \left[\frac{1}{4} (|\Delta|^2 + \epsilon_2^2-\epsilon_1^2({\bm k}))^2 + g^2(|\Delta|^2 +\epsilon_1^2({\bm k})+\epsilon_2^2 ) \right]^{1/2}
\end{eqnarray}
In the limit $\eta\gg |\mu|,T,g,|\Delta|$, we evaluate
\begin{eqnarray}
E^2_{-}({\bm k}) \approx \left(\frac{g^2}{\frac{k^2}{2m} + \eta} +\mu \right)^2 + |\Delta|^2 \left(1- \frac{2g^2}{(\frac{k^2}{2m} + \eta)^2}\right)  - \frac{2\mu g^4}{(\frac{k^2}{2m} + \eta)^3} 
\approx \left(\frac{g^2}{\frac{k^2}{2m} + \eta} +\mu \right)^2 + |\Delta|^2
\end{eqnarray}
and 
\begin{eqnarray}
E^2_{+}({\bm k}) \approx \left(\frac{k^2}{2m} + \eta -\mu \right)^2 +2 g^2 + \frac{2g^2}{(\frac{k^2}{2m} + \eta)^2}|\Delta|^2 \approx  \left(\frac{k^2}{2m} + \eta -\mu \right)^2=\epsilon_{1}^2({\bm k}) 
\end{eqnarray}

Generally, the Green function $\mathcal{G} $ can be written as
\begin{eqnarray}
\mathcal{G}(i\omega_n,{\bm k})= 
 \left( \begin{matrix}
 G(i\omega_n,{\bm k})  & F(i\omega_n,{\bm k}) \\
 F^{*}(i\omega_n,{\bm k})  &- G(-i\omega_n,{\bm k}) 
 \end{matrix}\right),
\end{eqnarray}
where $ G(i\omega_n,{\bm k})$ and $F(i\omega_n,{\bm k}) $ are the normal and anomalous Green functions. They are $2\times 2$ matrices acting on a two-band degree of freedom.
We denote the components of $ G(i\omega_n,{\bm k})$ as 
\begin{eqnarray}
G(i\omega_n,{\bm k}) = 
 \left( \begin{matrix}
 G_{\mathrm{c}}(i\omega_n,{\bm k}) & G_{\mathrm{cf}}(i\omega_n,{\bm k})\\
G_{\mathrm{fc}} (i\omega_n,{\bm k}) & G_{\mathrm{f}}(i\omega_n,{\bm k})
 \end{matrix}\right)
\end{eqnarray}
and similarly for $F(i\omega_n,{\bm k}) $. One obtains for $G(i\omega_n,{\bm k}) $
\begin{eqnarray}\nonumber
G(i\omega_n,{\bm k}) = 
\frac{-1}{ (\omega_n^2+E_{+}^2)(\omega_n^2+E_{-}^2)} \left( \begin{matrix}
(i\omega_n + \epsilon_1)(|\Delta|^2 + \epsilon_2^2+\omega_n^2) +g^2(i\omega_n-\epsilon_2)& g^3+g(i\omega_n+\epsilon_1)(i\omega_n+\epsilon_2)\\
g^3+g(i\omega_n+\epsilon_1)(i\omega_n+\epsilon_2) & (i\omega_n+\epsilon_2)(\omega_n^2+\epsilon_1^2) + g^2(i\omega_n-\epsilon_1)
 \end{matrix}\right)\\
\end{eqnarray}
and for $F(i\omega_n,{\bm k})$
\begin{eqnarray}
F(i\omega_n,{\bm k})= 
 \frac{|\Delta|}{ (\omega_n^2+E_{+}^2({\bm k}))(\omega_n^2+E_{-}^2({\bm k}))}\left( \begin{matrix}
g^2 & g(-i\omega_n-\epsilon_1({\bm k})) \\
g(i\omega_n-\epsilon_1({\bm k}))  & \omega_n^2+\epsilon^2_1
 \end{matrix}\right)
\end{eqnarray}
where 
\begin{eqnarray}
(\omega_n^2+E_{+}^2({\bm k}))(\omega_n^2+E_{-}^2({\bm k}))= g^4+2 g^2(\omega_n^2-\epsilon_1({\bm k})\epsilon_2) + (\omega_n^2+\epsilon_1^2({\bm k}))(\omega_n^2+|\Delta|^2+\epsilon_2^2)
\end{eqnarray}
At $\Delta =0 $, this simplifies to $F(i\omega_n,{\bm k})=0$ and 
\begin{eqnarray}
G(i\omega_n,{\bm k}) = 
\frac{-1}{(\omega_n+i\epsilon_1({\bm k}))(\omega_n+i\epsilon_2)+g^2} 
\left( \begin{matrix}
i\omega_n - \epsilon_2& g\\
g &i\omega_n - \epsilon_1({\bm k})
 \end{matrix}\right).
\end{eqnarray}
\subsubsection{Self-consistency equation}
The self-consistency equation for the gap amplitude is given by
\begin{align}
\frac{|\Delta|}{\lambda} =  |\Delta| T\sum_n\int_{k<K_{\mathrm{fb}}}  \frac{d{\bm k}}{(2\pi)^2} \frac{\omega_n^2+\epsilon^2_1({\bm k})}{(\omega_n^2+E_{+}^2({\bm k}))(\omega_n^2+E_{-}^2({\bm k}))}.
\end{align}
Let us consider some limiting cases. 
\\

1. At $g=0$, we recover the flat-band result Eq. (\ref{Trivial_SC}), namely
\begin{align}
\frac{|\Delta|}{\lambda} =  |\Delta| T\sum_n\int_{k<K_{\mathrm{fb}}}  \frac{d{\bm k}}{(2\pi)^2} \frac{1}{\omega_n^2+|\Delta|^2 + \mu^2},
\end{align}
which gives
\begin{align}
\frac{|\Delta|}{\lambda} = \frac{K_{\mathrm{fb}}^2}{4\pi} \frac{ |\Delta|}{2\sqrt{|\Delta|^2 + \mu^2}} \tanh\left(\frac{\sqrt{|\Delta|^2 + \mu^2}}{2T}\right).
\end{align}
\\

2. At $\mu=0$ but finite $g$, we obtain
\begin{align}
\frac{|\Delta|}{\lambda} =  \frac{m |\Delta| }{2\pi}  T\sum_n\int_{\eta}^{\eta+K^2_{\mathrm{fb}}/2m} dx \frac{\omega_n^2+x^2}{(\omega_n^2+g^2)^2 +\omega_n^2|\Delta|^2+ (\omega_n^2+|\Delta|^2)x^2},
\end{align}
For example, here expanding to the  lowest order in $g$ yields in the linearized regime
\begin{equation}
T_{\mathrm{PG}} = \frac{\lambda K^2_{\mathrm{fb}} }{16\pi} \left(1- \frac{4 m g^2}{\eta K^2_{\mathrm{fb}}} \right).
\end{equation}
The hybridization gives a small $m g^2/(\eta K^2_{\mathrm{fb}}) \ll 1$ correction to an effective interaction constant in Eq. (\ref{Trivial_SC}).
\\

3. In the limit when $\eta \gg g, T, |\mu|, |\Delta|$, we project to the flat-band as
\begin{align}
\frac{|\Delta|}{\lambda} =  |\Delta| T\sum_n\int_{k<K_{\mathrm{fb}}}  \frac{d{\bm k}}{(2\pi)^2} \frac{1}{\omega_n^2+\left(\frac{g^2}{k^2/(2m) + \eta} +\mu \right)^2 + |\Delta|^2},
\end{align}
In the linearized limit, we obtain Eq. (\ref{Sc_second}) of the main text. In the weak interaction regime, the text book expressions \cite{AGD} for the pairing amplitude are given by
\begin{align}
&|\Delta| = 2\pi T_{\mathrm{BCS}} \sqrt{\frac{2}{7\zeta(3)}}\sqrt{1-\frac{T}{T_{\mathrm{BCS}}}},~~~ T\rightarrow T_{\mathrm{BCS}},\\
& |\Delta| = |\Delta_0| -\sqrt{2\pi T |\Delta_0|}\left( 1- \frac{T}{8|\Delta_0|}\right)e^{-|\Delta_0|/T},~~~~ T\rightarrow 0,\\
& T_{\mathrm{BCS}} = \frac{\gamma}{\pi} \frac{k \mu^2 }{m g^2} \sqrt{K_{\mathrm{fb}}^2-k_{\mathrm{F}}^2} \exp\left\{-\frac{2\pi \mu^2}{\lambda m g^2}\left(1-\frac{\lambda}{\lambda_c} \right) \right\},~~~ |\Delta_0| = \frac{\pi}{\gamma} T_{\mathrm{BCS}},\\
&\frac{1}{\lambda_c} = \frac{K^2_{\mathrm{fb}} - 2 k^2_{\mathrm{F}}}{8\pi |\mu|}.
\end{align}
More generally for any strength of interaction, while at zero temperature keeping $|\Delta|$, we get
\begin{align}
\frac{|\Delta|}{\lambda} =  \frac{m |\Delta|}{4\pi } \int_{\eta}^{\eta+K^2_{\mathrm{fb}}/(2m)} \frac{ xd x}{\sqrt{\left(g^2 +\mu x \right)^2 +  x^2|\Delta|^2}}.
\end{align}
As a result 
\begin{align}\nonumber
&\frac{4\pi |\Delta|}{\lambda m} =  \frac{ |\Delta|}{\sqrt{\mu^2+|\Delta|^2}}
\bigg\{ \sqrt{\left(\frac{K^2_{\mathrm{fb}}}{2m}  +\eta + \frac{g^2\mu}{\mu^2+|\Delta|^2} \right)^2+ \frac{g^4|\Delta|^2}{(\mu^2+|\Delta|^2)^2}}
- \sqrt{\left(\eta + \frac{g^2\mu}{\mu^2+|\Delta|^2}\right)^2+ \frac{g^4|\Delta|^2}{(\mu^2+|\Delta|^2)^2}}\bigg\}
\\
&+ \frac{g^2\mu |\Delta|}{(\mu^2+|\Delta|^2)^{3/2}} \ln \left| \frac{\sqrt{\left(\frac{K^2_{\mathrm{fb}}}{2m}  +\eta + \frac{g^2\mu}{\mu^2+|\Delta|^2} \right)^2+ \frac{g^4|\Delta|^2}{(\mu^2+|\Delta|^2)^2}} -\frac{g^2\mu}{\mu^2+|\Delta|^2} - \frac{K^2_{\mathrm{fb}}}{2m}  - \eta}{\sqrt{\left(\eta + \frac{g^2\mu}{\mu^2+|\Delta|^2}\right)^2+ \frac{g^4|\Delta|^2}{(\mu^2+|\Delta|^2)^2}}  -\frac{g^2\mu}{\mu^2+|\Delta|^2} -\eta}\right|.
\end{align}
In the limit of small gap amplitude $|\mu| > |\Delta|$ and for the chemical potential $-g^2/\eta <\mu<0$, we obtain
\begin{equation}
\frac{1}{\lambda} = \frac{1}{\lambda_c} + \frac{m g^2}{2\pi \mu^2}\ln\left| \frac{\mu^2 k_{\mathrm{F}}}{m g^2 \Delta}  \sqrt{K_{\mathrm{fb}}^2-k_{\mathrm{F}}^2} \right|
\end{equation}
while in the strong interaction case when the chemical potential $\mu<-g^2/\eta $, the result is
\begin{equation}
\frac{ |\Delta|}{\lambda }= \frac{m |\Delta|}{4\pi \sqrt{|\Delta|^2+\mu^2} } \left\{
\frac{K_{\mathrm{fb}}^2}{2m} - \frac{|\Delta|^2 g^4}{2(|\Delta|^2+\mu^2)^2}\frac{1}{|\eta + \mu g^2/(|\Delta|^2+\mu^2) |} - 
\frac{\mu g^2}{|\Delta|^2+\mu^2}\ln \left|\frac{ K_{\mathrm{fb}}^2/(2m)}{\eta + \mu g^2/(|\Delta|^2+\mu^2)} \right|
\right\}
\end{equation}
When $g$-dependent terms considered as small corrections, one obtains
\begin{equation}
|\Delta| = \sqrt{\left[\frac{\lambda K_{\mathrm{fb}}^2}{8\pi }+ f(g, |\Delta|,\mu)\right]^2 -\mu^2}
\end{equation}
with $f(g, |\Delta|,\mu)$ for short hand notation. Specifically, at $\mu=0$, one obtains
\begin{equation}
\frac{ |\Delta|}{\lambda }= 
\frac{K_{\mathrm{fb}}^2}{8 \pi } - \frac{m g^4}{8 \pi \eta |\Delta|^2}
\end{equation}

The chemical potential shall be self-consistently determined through the particle number equation. We shall focus on the quasi-flat band only:
\begin{eqnarray}
N_{\mathrm{fb}} = \int_{k<K_{\mathrm{fb}}} \frac{d{\bm k}}{(2\pi)^2}\left\{1- \frac{\epsilon({\bm k})-\mu}{E({\bm k})}\tanh\left(\frac{E({\bm k})}{2T}\right)\right\}.
\end{eqnarray}
Here 
\begin{eqnarray}
E({\bm k})= \sqrt{\left(\frac{g^2}{\frac{k^2}{2m} + \eta} +\mu \right)^2 + |\Delta|^2}
\end{eqnarray}
Denote the filling factor as $\nu = 2 \pi N_{\mathrm{fb}}/K^2_{\mathrm{fb}}$. At zero temperature, we obtain
\begin{align}\nonumber
& (2\nu-1 )\frac{K^2_{\mathrm{fb}}}{2 m}= \frac{\mu}{\sqrt{\mu^2+|\Delta|^2}}
\bigg\{\sqrt{\left(\frac{K^2_{\mathrm{fb}}}{2m}  +\eta + \frac{g^2\mu}{\mu^2+|\Delta|^2} \right)^2+ \frac{g^4|\Delta|^2}{(\mu^2+|\Delta|^2)^2}}
- \sqrt{\left(\eta + \frac{g^2\mu}{\mu^2+|\Delta|^2}\right)^2+ \frac{g^4|\Delta|^2}{(\mu^2+|\Delta|^2)^2}}\bigg\}
\\
&- \frac{g^2|\Delta|^2}{(\mu^2+|\Delta|^2)^{3/2}} \ln \left| \frac{\sqrt{\left(\frac{K^2_{\mathrm{fb}}}{2m}  +\eta + \frac{g^2\mu}{\mu^2+|\Delta|^2} \right)^2+ \frac{g^4|\Delta|^2}{(\mu^2+|\Delta|^2)^2}} -\frac{g^2\mu}{\mu^2+|\Delta|^2} - \frac{K^2_{\mathrm{fb}}}{2m}  - \eta}{\sqrt{\left(\eta + \frac{g^2\mu}{\mu^2+|\Delta|^2}\right)^2+ \frac{g^4|\Delta|^2}{(\mu^2+|\Delta|^2)^2}}  -\frac{g^2\mu}{\mu^2+|\Delta|^2} -\eta}\right|.
\end{align}
For example, at $g=0$, one obtains equation for the chemical potential
\begin{align}
2\nu-1 = \frac{\mu}{\sqrt{\mu^2+|\Delta|^2}}.
\end{align}

\subsubsection{Supercurrent}
The current operator
\begin{equation}
\hat{{\bm J}}({\bm r}) = \frac{e}{m} \left( -i\boldsymbol{\partial}_r  - e{\bm A}({\bm r}) \tau_z \right)\frac{1+\sigma_z}{2},
\end{equation}
where ${\bm A}({\bm r})$ is the vector potential, $e<0$ is the electron charge, and $\tau_{z}$ is the Pauli matrix acting in Nambu space (in $c=1$ units).
The current density in the linear and static limits in Fourier components is given by  \cite{AGD}:
\begin{align}\nonumber\label{Appendix_current_main}
&{\bm J}({\bm q}) = -\frac{2e^2}{m}{\bm A}({\bm q}) \lim_{\tau \rightarrow +0} T\sum_n e^{i\omega_n \tau} \int  \frac{d{\bm k}}{(2\pi)^2}G_c(i\omega_n, {\bm k}) \\
&- \frac{2 e^2 }{m^2} \lim_{\tau \rightarrow +0} T\sum_n e^{i\omega_n \tau} \int  \frac{d{\bm k}}{(2\pi)^2} {\bm k}({\bm k} {\bm A}({\bm q}))
\bigg[ G_c(i\omega_n, {\bm k}_+)G_c(i\omega_n, {\bm k}_-)+F_c(i\omega_n, {\bm k}_+)F_c^{+}(i\omega_n, {\bm k}_-)\bigg],
\end{align}
where  ${\bm k}_{\pm} = {\bm k}\pm {\bm q}/2$, $\omega_n=(2n+1)\pi T$ is the Matsubara frequency with $n\in \mathcal{Z}$. 
The normal and anomalous Green functions are given by 
\begin{eqnarray}
G_c (i\omega_n, {\bm k})&=& 
-\frac{(i\omega_n + \epsilon_1({\bm k}))(|\Delta|^2 + \epsilon_2^2+\omega_n^2) +g^2(i\omega_n-\epsilon_2)}{(\omega_n^2+E_{+}^2({\bm k}))(\omega_n^2+E_{-}^2({\bm k})) },\\
F_c (i\omega_n, {\bm k}) &=&  \frac{g^2\Delta}{(\omega_n^2+E_{+}^2({\bm k}))(\omega_n^2+E_{-}^2({\bm k})) }.
\end{eqnarray}
The current density can be rewritten in components as
\begin{equation}
J_{\alpha}({\bm q}) = - \left(\frac{2e }{ \hbar}\right)^2Q_{\alpha \beta}({\bm q})  A_{\beta}({\bm q}).
\end{equation}
Here $Q_{\alpha \beta}({\bm q})$ is the phase stiffness (here $\hbar$ is restored for dimensionality).
Let us focus on an expression for the kernel at zero wave-vector. In this case, noting $Q_{\alpha \beta}(0) =Q \delta_{\alpha \beta}$, we obtain
\begin{align}\nonumber \label{current_definition_2}
&Q = -\frac{1}{2m}  \lim_{\tau \rightarrow +0} T\sum_n e^{i\omega_n \tau} \int_{0}^{\infty} 
\frac{k dk}{2\pi} \frac{
(i\omega_n + \epsilon_1({\bm k}))(|\Delta|^2 + \epsilon_2^2+\omega_n^2) +g^2(i\omega_n-\epsilon_2)
}{g^4+2 g^2(\omega_n^2-\epsilon_1({\bm k})\epsilon_2) + (\omega_n^2+\epsilon_1^2({\bm k}))(\omega_n^2+|\Delta|^2+\epsilon_2^2) }\\
&+\frac{1}{4m^2 } \lim_{\tau \rightarrow +0} T\sum_n e^{i\omega_n \tau}
\int_{0}^{\infty} \frac{k^3 dk}{2\pi} \frac{|\Delta|^2 g^4 +\left[(i\omega_n +\epsilon_1({\bm k})) (\omega_n^2+\epsilon_2^2+ |\Delta|^2)+g^2(i\omega_n-\epsilon_2) \right]^2 }{[g^4+2 g^2(\omega_n^2-\epsilon_1({\bm k})\epsilon_2) + (\omega_n^2+\epsilon_1^2({\bm k}))(\omega_n^2+|\Delta|^2+\epsilon_2^2) ]^2}.
\end{align}
Integrating in the second term in (\ref{current_definition_2})  by parts over $k$ results in
\begin{align}\nonumber
&
\frac{1}{4m^2}\int_{0}^{\infty} \frac{k^3 dk}{2\pi} \frac{\left[(i\omega_n +\epsilon_1({\bm k})) (\omega_n^2+\epsilon_2^2+ |\Delta|^2)+g^2(i\omega_n-\epsilon_2) \right]^2-|\Delta|^2 g^4 }{[g^4+2 g^2(\omega_n^2-\epsilon_1({\bm k})\epsilon_2) + (\omega_n^2+\epsilon_1^2({\bm k}))(\omega_n^2+|\Delta|^2+\epsilon_2^2) ]^2} 
\\\nonumber
&=- \frac{k^2}{8\pi m} \frac{(i\omega_n + \epsilon_1({\bm k}))(|\Delta|^2 + \epsilon_2^2+\omega_n^2) +g^2(i\omega_n-\epsilon_2)}{g^4+2 g^2(\omega_n^2-\epsilon_1({\bm k})\epsilon_2) + (\omega_n^2+\epsilon_1^2({\bm k}))(\omega_n^2+|\Delta|^2+\epsilon_2^2)} \bigg|_{0}^{\infty} 
\\
&+ \int_{0}^{\infty} \frac{k dk}{4\pi m} \frac{(i\omega_n + \epsilon_1({\bm k}))(|\Delta|^2 + \epsilon_2^2+\omega_n^2) +g^2(i\omega_n-\epsilon_2)}{g^4+2 g^2(\omega_n^2-\epsilon_1({\bm k})\epsilon_2) + (\omega_n^2+\epsilon_1^2({\bm k}))(\omega_n^2+|\Delta|^2+\epsilon_2^2)} 
\end{align}
Here, using $\lim_{\tau \rightarrow +0} T\sum_n e^{i\omega_n \tau}  \frac{1}{i\omega_n - k^2/(2m)} = \frac{1}{1+\exp[k^2/(2m T)]}$ for the $k\rightarrow \infty$ limit, we obtain
\begin{eqnarray}\label{Q_after_parts}
Q =
 \frac{T  |\Delta|^2 g^4}{4\pi m^2 }\sum_n \int_{0}^{\infty}  \frac{k^3 dk}{(\omega_n^2+E_{+}^2({\bm k}))^2(\omega_n^2+E_{-}^2({\bm k}))^2}.
\end{eqnarray}
Thus the lowest order in $g$ it gives
\begin{eqnarray}
Q=  \frac{T|\Delta|^2 g^4}{4\pi m^2 }\sum_n \int_{0}^{\infty}  \frac{k^3 dk}{(\omega_n^2+\epsilon_1^2({\bm k}))^2(\omega_n^2+|\Delta|^2+\epsilon_2^2)^2}.
\end{eqnarray}
So that we obtain (\ref{Q_strange}) of the main text. On the other hand, projecting in (\ref{Q_after_parts}) to the flat-band yields
\begin{equation}
Q =
\frac{T  |\Delta|^2 g^4}{4\pi m^2}\sum_n \int_{0}^{\infty} \frac{1}{\left(\frac{k^2}{2m}+\eta \right)^4} \frac{k^3 dk}{\left[\omega_n^2+|\Delta|^2 + \left(\frac{g^2}{k^2/(2m) + \eta} +\mu \right)^2  \right]^2}.
\end{equation}
which leads to (\ref{Q_main_2})
\begin{eqnarray}
Q = \frac{k_{\mathrm{F}}^2\mu^2}{8 m g^2} T\sum_n  \frac{|\Delta|^2}{(\omega_n^2+|\Delta|^2)^{3/2}}.
\end{eqnarray} 
Taking into account the wave-vector dependence in $Q ({\bm q})$, a straightforward calculations
in the London limit, at zero temperature, yields
\begin{equation}
Q ({\bm q})  = \frac{k_{\mathrm{F}}^2\mu^2}{8\pi m g^2} \left(1 - \frac{v_{\mathrm{F}}^2q^2}{8|\Delta|^2}\right).
\end{equation}
While in the Pippard limiting case, $v_{\mathrm{F}} q \gg |\Delta|$, at zero temperature we obtain
\begin{equation}
Q({\bm q}) = \frac{k_{\mathrm{F}}^2\mu^2}{ m g^2}   \frac{|\Delta|^2}{v_{\mathrm{F}}^2 q^2}\ln\left| \frac{v_{\mathrm{F}} q}{\Delta} \right|.
\end{equation}

\subsection{Two-particle bound states in the two-band model}
Consider the eigenvalue equation for the problem of two non-interacting particles:
\begin{align}\nonumber
&\left\{\frac{1+\sigma_z}{2}\left(-\frac{\boldsymbol{\nabla}_{{\bm r}_1}^2}{2m} +\eta \right) + g\sigma_x \right\}_{\alpha, \nu} \Psi_{\nu,\beta}({\bm r}_1, {\bm r}_2) \\
&+ \left\{\frac{1+\sigma_z}{2}\left(-\frac{\boldsymbol{\nabla}_{{\bm r}_2}^2}{2m} +\eta \right) + g\sigma_x \right\}_{\beta, \nu} \Psi_{\alpha,\nu}({\bm r}_1, {\bm r}_2)= E \Psi_{\alpha,\beta}({\bm r}_1, {\bm r}_2)
\end{align}
The wave-function symmetric with respect to band index has the form
\begin{eqnarray}
\Psi_{\alpha,\beta}({\bm r}_1, {\bm r}_2) = \left(
 \begin{matrix}
a({\bm r}_1, {\bm r}_2)& b({\bm r}_1, {\bm r}_2)\\
b({\bm r}_1, {\bm r}_2) &c({\bm r}_1, {\bm r}_2)
 \end{matrix}\right)_{\alpha,\beta}
\end{eqnarray}
Consider coordinate symmetrization $\boldsymbol{\rho} = {\bm r}_1-{\bm r}_2$ and ${\bm R} = \frac{{\bm r}_1+{\bm r}_2}{2}$, so that
\begin{equation}
\Psi_{\alpha,\beta}({\bm r}_1, {\bm r}_2) \rightarrow \Psi_{\alpha,\beta}(\boldsymbol{\rho} , {\bm R})
\end{equation}
Let us perform Fourier transformation with respect to ${\bm R}$ with $\exp(i {\bm Q \bm R})$ and then set $Q=0$. As a result $\boldsymbol{\nabla}_{{\bm r}_{1,2}}^2\rightarrow \boldsymbol{\nabla}_{\boldsymbol{\rho}}^2$ and
\begin{eqnarray}
\Psi_{\alpha,\beta}({\bm \rho}) = \left(
 \begin{matrix}
c({\bm \rho})& b({\bm \rho}) \\
b({\bm \rho}) &d({\bm \rho})
 \end{matrix}\right)_{\alpha,\beta}
\end{eqnarray}

It is convenient to  consider the finite range interaction potential in the form $$U({\bm \rho}) = -U_0 \theta (a- \rho),~~ U_0>0.$$ We obtain
\begin{align}
&\left(-\frac{\boldsymbol{\nabla}_{{\bm \rho}}^2}{2m} +\eta \right) \left(
 \begin{matrix}
2c& b \\
b&0
 \end{matrix}\right)  + g \left(
 \begin{matrix}
2b& c+d \\
c+d& 2b
 \end{matrix}\right)  = (E-U({\bm \rho})) \left(
 \begin{matrix}
c& b \\
b&d
 \end{matrix}\right)
\end{align}
We first find a solution for $$d = \frac{2 g b}{E- U({\bm \rho})},$$ while a system of equations for two other components $(c,b)$ is given by
\begin{align}
& 
\left(
 \begin{matrix}
-\frac{\boldsymbol{\nabla}_{{\bm \rho}}^2}{2m} +\eta -\frac{E- U({\bm \rho})}{2}& g \\
g& -\frac{\boldsymbol{\nabla}_{{\bm \rho}}^2}{2m} +\eta -E+ U({\bm \rho}) + \frac{2 g^2}{E- U({\bm \rho})}
 \end{matrix}\right)  
 \left(
 \begin{matrix}
c\\
b
 \end{matrix}\right)  
 = 0
\end{align}

In a two-band system we expect three types of bound states. Hence, we seek a solution for a system of linear equations for a vector $$\Lambda_{1,2,3}=  \left(
 \begin{matrix}
c_{1,2,3}\\
b_{1,2,3}
 \end{matrix}\right)  $$ For the basis let us take homogeneous solutions of the system at $U=0$:
 \begin{align}
& 
\left(
 \begin{matrix}
\eta -\frac{E(0)}{2}& g \\
g& \eta -E(0) + \frac{2 g^2}{E(0)}
 \end{matrix}\right)  
 \left(
 \begin{matrix}
c\\
b
 \end{matrix}\right)  
 = 0
\end{align}
The solutions are
\begin{align}
&E_1(0) = \eta >0\\
 &E_2(0) = \eta+\sqrt{\eta + 4 g^2} \approx 2\eta>0\\
  &E_3(0) = \eta-\sqrt{\eta + 4 g^2}\approx -2g^2/\eta <0
\end{align} 

Taking into account the interaction potential $U$, the energies will be shifted as
$$ E_i = E_i(0) + \epsilon_i, ~~~ i=1,2,3$$

Further one may follow the solution described in Landau-Lifshitz, volume 3, for a situation when $|E_i(0)|>U_0$. We seek for a solution in the form
$$\Psi_i = \Lambda_i f_i(\rho).$$ At $\rho<a$, we neglect exponentially small $\epsilon_i$, while at $\rho>a$, we keep it.

At $\rho<a$, function $f_i(\rho)$ satisfies an equation
 \begin{align}
\left[-\frac{\boldsymbol{\nabla}_{{\bm \rho}}^2}{2m} + 
\Lambda_i^T \left(
 \begin{matrix}
 -\frac{U_0}{2} & 0 \\
0&  -U_0 (1+ \frac{2 g^2}{E^2_i(0)})
 \end{matrix}\right)  \Lambda_i
\right]  f_i(\rho)  
 = 0
\end{align}
At $a<\rho$, function $f_i(\rho)$ satisfies an equation
 \begin{align}
\left[-\frac{\boldsymbol{\nabla}_{{\bm \rho}}^2}{2m} + 
\Lambda_i^T \left(
 \begin{matrix}
 -\frac{\epsilon_i}{2} & 0 \\
0&  -\epsilon_i (1+ \frac{2 g^2}{E^2_i(0)})
 \end{matrix}\right)  \Lambda_i
\right]  f_i(\rho)  
 = 0
\end{align}
We find the two-particle bound energies
\begin{align}
&\epsilon_1 = - \frac{2}{m a^2} \exp\left( - \frac{2}{m a^2 U_0}\right)\\
&\epsilon_2 = - \frac{4}{m a^2} \exp\left( - \frac{4}{m a^2 U_0}\right),\\
&\epsilon_3 = - \frac{2}{m a^2} \frac{g^2}{\eta^2}\exp\left( - \frac{2}{m a^2 U_0}\frac{g^2}{\eta^2} \right), ~~~ \frac{2 g^2}{\eta}> U_0
\end{align}
where we assume $\eta\gg g$. Also $\Lambda_1^T=(-2g,\eta)/\sqrt{\eta^2+4 g^2}$, $\Lambda_2^T=(-2g,\eta - \sqrt{\eta^2+4g^2})/\sqrt{4g^2+(\eta - \sqrt{\eta^2+4g^2})^2} \approx(1,0)$, $\Lambda_3^T =(-2g,\eta + \sqrt{\eta^2+4g^2})/\sqrt{4g^2+(\eta + \sqrt{\eta^2+4g^2})^2} \approx (0,1)$.

\subsection{\label{sec:Appendix_B}Dynamics at low temperatures}
Let us consider the dynamics of the pairing function in the isolated flat-band case, taking into account small deviations around the saddle-point solution $$\Delta_j(\tau) = [|\Delta|+\delta \Delta(\tau)] e^{i\theta_j(\tau)}.$$ 
Writing an effective action $S$ as a sum over all domains $S = \sum_j S_j$, neglecting the electromagnetic field term, one obtains, \cite{Golubev_action}
\begin{eqnarray}\label{Effective_action}
S_j=- \mathrm{Tr}\ln\left[-\partial_\tau + (\mu+ie\tilde{V_j})\tau_z + (|\Delta|+\delta \Delta) \tau_x \right]
+(4\pi/K^2_{\mathrm{fb}})\int_0^{\frac{1}{T}} d\tau\left[ ien_i \tilde{V_j} +(|\Delta|+\delta \Delta)^2/\lambda\right],
\end{eqnarray}
where $en_i$ denotes the background charge density of the ions. 
The electromagnetic potential and superconducting phase are expressed in a gauge-invariant form: $\tilde{V}_j = V - \partial_\tau \theta_j/(2e)$. 

Expanding the action (\ref{Effective_action}) to quadratic order in small $e\tilde{V}$ and $\delta \Delta$ around the saddle-point solution of Eq. (\ref{Trivial_SC}) gives
$S= \sum_j(S_j +\delta S_j)$, where now $S_j$ is given by (\ref{toy_action}) in the main text, and $\delta S_j$ represents the fluctuation part of the action:
\begin{eqnarray}\label{deltaS0}
\delta S_j= \int_0^{\frac{1}{T}} d\tau
\left\{
2i \gamma_{\Delta V} e\tilde{V_j}
+ \gamma_V (e\tilde{V_j})^2 +\gamma_{\Delta} \delta \Delta^2
\right\},
\end{eqnarray}
where the coefficients $\gamma_{\Delta}$ and $\gamma_{V}$ are given by 
$$ \gamma_{\Delta} = - \frac{|\Delta|^2}{2 E_0} \frac{\partial}{\partial E_0} \left[ \frac{1}{E_0}\tanh\left(\frac{E_0}{2T}\right)\right]$$ and  $$ \gamma_{V} = \gamma_{\Delta} + \frac{1}{2} \frac{\partial}{\partial E_0}\tanh\left(\frac{E_0}{2T}\right)$$ 
with $E_0= \sqrt{\mu^2+|\Delta|^2}$. At low temperatures ($T\ll T_{\mathrm{PG}}$), both coefficients simplify to 
\begin{equation}
\gamma_{\Delta} \equiv \gamma_{V} =\frac{ (1-\nu)\nu}{T_{\mathrm{PG}}}.
\end{equation}
The first term in $\delta S_j$ arises only when the particle-hole symmetry is broken. We obtain
\begin{equation}
\gamma_{\Delta V} = \nu - \nu|_{|\Delta|\rightarrow |\Delta|+\delta\Delta} \approx \mathrm{sign}(2\nu-1) \frac{\delta\Delta}{4T_{\mathrm{PG}}}.
\end{equation}
This term mediates a coupling between phase and amplitude fluctuations, which becomes stronger as the system deviates further from the particle-hole symmetric state.
We note that the expansion breaks down near the edges of the band, where $|\Delta|$ vanishes.
Integrating out the Gaussian fluctuations of $\delta\Delta$ and replacing $\theta_j(\tau) \rightarrow \theta_j(\tau) + e \int_0^{\tau} V_j(\tau') d\tau'$, we obtain the fluctuation
\begin{equation}
\delta S_j= \int_0^{\frac{1}{T}} d\tau \frac{(\partial_\tau \theta_j)^2}{8E_0} \mathrm{tanh}\left(\frac{E_0}{2T}\right).
\end{equation}
Using the expression for the filling-factor and (\ref{mu_main}), we can express the action for phase fluctuations as
\begin{equation}\label{deltaS1}
\delta S_j= \int_0^{\frac{1}{T}} d\tau \frac{(\partial_\tau \theta_j)^2}{16T_{\mathrm{PG}}}.
\end{equation}
Interestingly, the coefficient in this expression does not depend on the filling factor of the flat band after accounting for amplitude fluctuations.
Following \cite{Efetov_granular}, the phase correlator on domain $j$ is given by 
\begin{equation}\label{Correlator}
\langle \exp\{i [ \theta_j(0)- \theta_j(\tau)]\}\rangle   = \exp\left(- 4 \tau T_{\mathrm{PG}}\right) 
\end{equation}
at zero temperature. The correlator approaches unity at the crossover temperature $T\rightarrow T_{\mathrm{PG}}$.
In real time $(\tau \rightarrow i t)$, the phase correlations oscillate rapidly, with a period proportional to $\propto T^{-1}_{\mathrm{PG}}$. 

\end{document}